\documentclass[journal, onecolumn, 12pt, draftclsnofoot]{IEEEtran}

\newtheorem{thm}{Theorem}

\newtheorem{remk}{Remark}

\usepackage[final]{graphicx}
\usepackage[reqno]{amsmath}
\usepackage{amssymb}
\usepackage{amsmath}
\usepackage{subfig}
\usepackage{epstopdf}
\usepackage[paper=letterpaper,tmargin=1in,bmargin=.8in,left=.8in, right=.8in]{geometry}
\usepackage{algorithm}
\usepackage{algorithmicx}
\usepackage{algpseudocode}

\begin{document}

\sloppy
 
\title{A Blockchain Example for Cooperative Interference Management}
\author{{\large{Aly El Gamal, {\em Member, IEEE}, and Hesham El Gamal, {\em Fellow, IEEE}}}
\thanks{Aly El Gamal is with the ECE Department of Purdue University, West Lafayette, IN (e-mail: elgamala@purdue.edu). Hesham El Gamal is with the ECE Department of the Ohio State University, Columbus, OH (e-mail: helgamal@ece.osu.edu)}}

\maketitle

\begin{abstract}
We present an example where a distributed coordinated protocol supported by a blockchain-enabled monetary mechanism leads to achieving optimal information theoretic degrees of freedom gains. The considered setting is that of a linear interference network, where cooperative transmission is allowed, but at no cost in terms of the overall backhaul load. In other words, the average number of messages assigned to a transmitter is one. We show that a simple monetary mechanism that consists only of one coin type can enable the achievability of the optimal centralized solution. The proposed greedy distributed algorithm relies on incentivizing the users to share their resources in one channel use, in return of credit they receive for maximizing their rate gains in future channel uses. This example is the first in its class and it opens the door for constructing a unified framework for blockchain-enabled monetary mechanisms for optimal interference management and spectrum sharing. 
\end{abstract}

\begin{IEEEkeywords}
Blockchain, CoMP, interference management, monetary mechanism, backhaul load.
\end{IEEEkeywords}
\section{Introduction}
The problem of interference management in next generation infrastructural wireless networks is expected to have a very different nature from its current one. On one hand, the anticipated quality of service requirements from these networks are unprecedented, with new data-intensive applications like virtual reality and 360 videos as well as new applications that have very high reliability and low latency requirements like those supported by vehicular networks. 
On the other hand, new technological advancements are opening opportunities to meet the new requirements and achieve significant performance gains. In particular, cooperative communication (also known as \textbf{Coordinated Multi-Point} or CoMP) has been shown to hold the promise of delivering significant rate gains while simultaneously reducing the delay requirements, even when restricted to local cooperation strategies that do not require sharing of message between distant nodes in the network (see e.g.,~\cite{ElGamal-Veeravalli-Cambridge2018} and \cite{CoMP-book}). The potential of cooperative communication, however, seems to be limited thus far to the centralized scenario, due to the fact that the \textbf{achieved gains are asymmetric} across all participating users in the network in every channel use (see e.g.,~\cite{ElGamal-Annapureddy-Veeravalli-IT14}). Having a \textbf{robust economic framework} could alleviate this problem, and enable efficient distributed cooperation strategies, where \textbf{mutual trust is established by a monetary mechanism} that incentivizes users to offer help in one channel use in exchane of credit that they could use in future channel uses. Here, we present the first example of such a technique {\bf blockchain empowered} technique. 

In this work, we consider the linear interference network introduced by Wyner~\cite{Wyner}. A message can be assigned to more than one transmitter to enable CoMP transmission, but the average number of messages per transmitter has to be at most one. The considered rate criterion is the per user Degrees of Freedom (puDoF). This problem was settled in~\cite{ElGamal-Veeravalli-ISIT14} for the case when messages are assigned to transmitters through a centralized controller. The \textbf{optimal solution reaches a non-integer puDoF, where some users have to be inactive} in each channel use. Hence, a myopic solution where each user optimizes its own rate in each channel use is suboptimal. Here, we show how a greedy distributed algorithm that maximizes local benefit leads to the optimal centralized solution, when supported by a simple monetary mechanism that has only one coin type and a simple coordination protocol that relies on $2$-bit messages.  We believe that the presented example \textbf{opens the door for a general paradigm of backhaul and spectrum resource sharing based on blockchain-enabled monetary mechanisms}.
\section{System Model and Notation}\label{sec:systemmodel}
We use the standard model for the $K$-user interference channel with single-antenna transmitters and receivers,
\begin{equation}
Y_i(t) = \sum_{j=1}^{K} H_{i,j}(t) X_j(t) + Z_i(t),
\end{equation}
where $t$ is the time index, $X_j(t)$ is the transmitted signal of transmitter $j$, $Y_i(t)$ is the received signal at receiver $i$, $Z_i(t)$ is the zero mean unit variance Gaussian noise at receiver $i$, and $H_{i,j}(t)$ is the channel coefficient from transmitter $j$ to receiver $i$ over the time slot $t$. We remove the time index in the rest of the paper for brevity unless it is needed. 
Finally, we use $[K]$ to denote the set $\{1,2,\ldots,K\}$.

\subsection{Channel Model}
Each transmitter is connected to its corresponding receiver as well as one following receiver, and the last transmitter is only connected to its corresponding receiver. More precisely,

\begin{equation}\label{eq:channel}
H_{i,j} = 0 \text { iff } i \notin \{j,j+1\},\forall i,j \in [K],
\end{equation}
and all non-zero channel coefficients are drawn from a continuous joint distribution. Finally, we assume that global channel state information is available at all nodes. 

\subsection{Message Assignment}
For each $i \in [K]$, let $W_i$ be the message intended for receiver $i$, and ${\cal T}_i \subseteq [K]$ be the transmit set of receiver $i$, i.e., those transmitters with the knowledge of $W_i$. The transmitters in ${\cal T}_i$ cooperatively transmit the message $W_i$ to the receiver $i$. Although cooperative transmission is allowed, the average transmit set size is upper bounded by a unity backhaul load constraint,
\begin{equation}\label{eq:backhaul_constraint}
\frac{\sum_{i=1}^K |{\cal T}_i|}{K} \leq 1.
\end{equation}

\subsection{Message Assignment Strategy}

A message assignment strategy is defined by a sequence of supersets. The $k^{th}$ element in the sequence consists of the transmit sets for a $k$-user channel. We use message assignment strategies to define a pattern for assigning messages to transmitters in large networks. For brevity, we refer to a message assignment strategy by a message assignment, when it is clear from context how to apply the assignment to an arbitrarily large network.



\subsection{Degrees of Freedom}
Let $P$ be the average transmit power constraint at each transmitter, and let ${\cal W}_i$ denote the alphabet for message $W_i$. Then the rates $R_i(P) = \frac{\log|{\cal W}_i|}{n}$ are achievable if the decoding error probabilities of all messages can be simultaneously made arbitrarily small for a large enough coding block length $n$, and this holds for almost all channel realizations. The degrees of freedom $d_i, i\in[K],$ are defined as $d_i=\lim_{P \rightarrow \infty} \frac{R_i(P)}{\log P}$. The DoF region ${\cal D}$ is the closure of the set of all achievable DoF tuples. The total number of degrees of freedom ($\eta$) is the maximum value of the sum of the achievable degrees of freedom, $\eta=\max_{\cal D} \sum_{i \in [K]} d_i$.

For a $K$-user channel, we define $\eta(K)$ as the best achievable $\eta$ over all choices of transmit sets satisfying the backhaul load constraint in \eqref{eq:backhaul_constraint}. 
In order to simplify our analysis, we define the asymptotic per user DoF $\tau$ to measure how $\eta(K)$ scales with $K$ while all other parameters are fixed,
\begin{equation}
\tau = \lim_{K\rightarrow \infty} \frac{\eta(K)}{K},
\end{equation}

We call a message assignment strategy \emph{optimal} for a sequence of $K$-user channels, $K\in\{1,2,\ldots\}$,  if and only if there exists a sequence of coding schemes achieving $\tau$ using the transmit sets defined by the message assignment strategy. 
\section{Prior Work: Centralized Message Assignment}
\begin{figure}[htb]
\centering
\includegraphics[width=0.4\columnwidth]{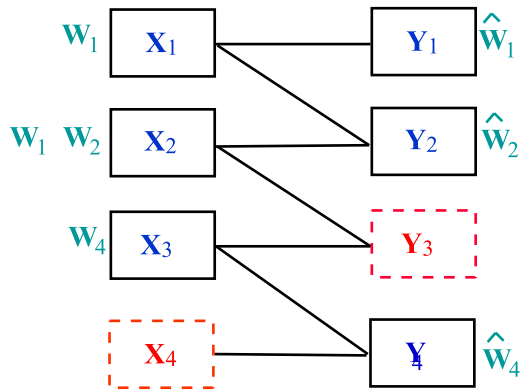}
\caption{Degrees of Freedom-optimal scheme when messages can be assigned through a centralized controller.}
\label{fig:bone}
\end{figure}

The considered setting was studied in~\cite{ElGamal-Veeravalli-ISIT14}, while assuming that messages are assigned to transmitters through a centralized controller that is aware of the global network topology. It was shown that $\tau=\frac{3}{4}$, and is achievable by splitting the network into subnetworks; each consisting of four transmitter-receiver pairs, and using the message assignment illustrated in Figure~\ref{fig:bone} in each subnetwork. Only signals corresponding to the first subnetwork in a general $K$-user network are shown in the figure. The signals in the dashed boxes are deactivated. Note that the last transmitter in each subnetwork is deactivated to eliminate inter-subnetwork interference. Also, exactly four message instances are assigned for each group of four messages, which is consistent with the constraint in~\eqref{eq:backhaul_constraint}. We also note that the third message in each subnetwork is inactive, which indicates the necessity of an incentive for the corresponding users for a distributed algorithm to reach the optimal solution. For further explanation of the transmission scheme, see~\cite[Chapter $7$]{ElGamal-Veeravalli-Cambridge2018}.

\section{Blockchain-Based Distributed Scheme}
In this section, we demonstrate how a greedy distributed algorithm can reach the optimal centralized assignment of messages and transmission scheme, when supported by appropriate incentives through a blockchain-based monetary mechanism. We start with the simple assumption that the $i^{\text{th}}$ user has a transmitter-receiver pair index $i$, and is responsible for delivering $W_i$. 
\subsection{Monetary Mechanism}
Consider the following simple example of a monetary mechanism: There is only one coin type, and this coin could be given from a node (transmitter-receiver pair) to another for one of the following reasons:
\begin{enumerate}
\item The payer's transmitter will cause interference at the payee's receiver.

\item The payer will rent the payee's transmitter.




\end{enumerate}
The underlying concept is based on an interference avoidance transmission strategy, where a node has to \textbf{pay in order to be active}. A paid coin goes either to the following node as in the first case above, or to the preceding node as in the second case above. 
\subsection{Coordination Protocol}
Besides the monetary mechanism, the transmission strategy relies on a coordination protocol, where a node (say with index $i$) indicates to the following node in the network (with index $i+1$) one of the following scenarios:
\begin{itemize}
\item Coordination Message $1$ ($CM1$): Node $i$ is willing to get paid by node $i+1$ to turn off transmitter $i$.
\item $CM2$: Node $i$ is willing to get paid by node $i+1$ to use transmitter $i$.
\item $CM3$: Node $i$ is not willing to get paid by node $i+1$ and node $i+1$ can be active (it is possible for receiver $i+1$ to observe no interference). 
\item $CM4$: Node $i$ is not willing to get paid by node $i+1$ and node $i+1$ cannot be active, i.e., its receiver will observe interference that cannot be eliminated, because otherwise the overall backhaul load constraint would be violated.
\end{itemize}
We have the following remarks on the extension of the above protocol for general network topologies.

\begin{remk}
For $CM2$, transmitter $i$ can only deliver messages $W_i$ or $W_{i+1}$ in a linear network. In a general setting, node $i$ can open an auction among all nodes interested in using its transmitter. However, this may come at an extra cost in terms of transaction overhead and delay.
\end{remk}
\begin{remk} For $CM3$ and $CM4$, the control information about the possibility of having receiver $i+1$ decode its desired message comes only from node $i$, since in the considered linear network topology, transmitter $i$ is the only source for interference at receiver $i+1$. In a general setting, such information will be aggregated at node $i+1$ from coordination messages received from all nodes that have transmitters interfering at receiver $i+1$.
\end{remk}

\subsection{Greedy Algorithm}
We now explain a greedy distributed algorithm  that \textbf{reaches the optimal centralized solution} for the message assignment and transmission strategy. The algorithm proceeds in ascending order of the node index, and each node decides first to be active if possible, and then maximizes its monetary credit. Both decision criteria are considered in a greedy myopic fashion. The options available for each node are based on the coordination message it receives from the preceding node. The new coordination message passed by a node to its following node, is based on \textbf{awareness of the backhaul constraint and network topology}. We provide Algorithm~\ref{alg:the_alg} below, and then show how it reaches the solution illustrated in Figure~\ref{fig:bone} that is proved to be the optimal centralized solution in~\cite{ElGamal-Veeravalli-ISIT14}. 

\begin{algorithm}
\caption{Greedy algorithm for blockchain-based cooperative interference management in linear networks.}
\begin{algorithmic}[1]
    \If {Received $CM4$ from node $i-1$}
    \State{Send $CM2$ to node $i+1$.}
    \State{Activate a subroutine that gives node $i+1$ access to transmitter $i$ if a coin is received.}
    \ElsIf {Received $CM3$ from node $i-1$}
    \State{Transmit $W_i$ from transmitter $i$.}
    \State{Pay node $i+1$.}
    \If{Node $i-1$ paid node $i$}
    \State{Design the transmit beam for $W_{i-1}$ at transmitter $i$ to cancel its interference at receiver $i$;}
    \State{Send $CM4$ to node $i+1$;}
    \Else{} 
    \State{Send $CM3$ to node $i+1$;}
    \EndIf
\ElsIf {Received $CM2$ from node $i-1$}
    \State{Pay node $i-1$ and transmit $W_i$ from transmitter $i-1$;}
    \State{Send $CM1$ to node $i+1$.}
    \State{Activate a subroutine that enables node $i+1$ to turn off transmitter $i$ if a coin is received.}
    \Else{}
    \State{Pay node $i-1$ and turn off transmitter $i-1$;}
    \State{Transmit $W_i$ from transmitter $i$;}
    \State{Pay node $i+1$.}
    \State{Send $CM3$ to node $i+1$;}
    \EndIf
\end{algorithmic}  
\label{alg:the_alg}
\end{algorithm}
\begin{thm}
Applying Algorithm $1$ in an ascending order of index, the optimal message assignment is reached, and the optimal per user DoF $\tau=\frac{3}{4}$ is achieved.
\end{thm}
\begin{IEEEproof}
We first trace Algorithm~\ref{alg:the_alg} for the first $4$ users in the network, and verify that we get the solution illustrated in Figure~\ref{fig:bone}. The execution of the algorithm at the first four users will be as follows.
\begin{enumerate}
\item We assume that the first user has an initial coordination message indicating that the preceding node is not willing to get paid but the first user can be active. Following lines 5-6 and 11 of the algorithm, $W_1$ will be transmitted from transmitter $1$, node $1$ will send a coordination message to node $2$ indicating that it is not willing to get paid, but receiver $2$ can be active, and finally node $1$ will pay node $2$ because transmitter $1$ will cause interference at receiver $2$. 

\item Based on the received coordination message, the second node will execute lines  5-6 and 8-9 of the algorithm. Hence, both $W_1$ and $W_2$ will be transmitted from transmitter $2$ for interference cancellation and message delivery, respectively. Also, a coordination message will be sent to node $3$ indicating that receiver $3$ has to be inactive, and a coin will be paid to node $3$. 

\item Based on the received coordination message, the third node will execute lines 2-3 of the algorithm and send a coordination message to node $4$ indicating that node $3$ is willing to rent out its transmitter for a coin.

\item Node $4$ would then execute lines 14-16 of the algorithm, by paying node $3$ to use transmitter $3$ for delivering $W_4$, and sending a coordination message to node $5$ indicating that it can get paid to turn off transmitter $4$.
\end{enumerate}
We note that node $5$ would execute lines 18-21, which form a set of actions analogous to what node $1$ executed, except for paying the node above to turn off its transmitter. If we follow the rest of the network, we would find that all nodes that have the same index modulo $4$ would execute a similar set of actions. The theorem statement hence follows, as the resulting message assignment and transmission scheme are identical to the optimal one found in~\cite{ElGamal-Veeravalli-ISIT14}, and $3$ DoF is achieved for each subnetwork of $4$ users. 
\end{IEEEproof}
We note that the coordination message from the second to the third node indicating that receiver $3$ has to be inactive, is based on the information-theoretic analysis in~\cite{ElGamal-Veeravalli-ISIT14} where it is shown that optimal assignment of messages to transmitters in a linear network that satisfy an average of one message per transmitter, never assigns one message to more than two transmitters. Hence, having downloaded $W_1$ for interference cancellation, node $2$ knows that $W_1$ would interfere at receiver $3$, and the optimal solution would not be reached if transmitter $3$ downloads the same message for interference cancellation. Node $2$ then tells node $3$ that its receiver has to be inactive and pays a coin for it. In general, this is the justification for having coordination message $CM4$. 

\section{Discussion}\label{sec:discussion}
We discuss below different aspects of the presented example, and use Algorithm~\ref{alg:the_alg} as a reference to illustrate key concepts regarding the support of distributed schemes for cooperative interference management with blockchain-based monetary incentives.

\subsection{Coin Allocation:} In the example illustrated above, each node with an index $i$ such that $i \text{ mod } 4=1$ will pay two coins, one to the preceding node and another to the following node - with the exception of the first node in the network that will be paying only the following node - and each node with an index $i$ such that $i \text{ mod }4=3$ will receive two coins, one from the preceding node and another from the following node. Every other node (with an even index) will pay one coin and receive one coin. Hence, overall we need $\frac{K}{2}$ coins in a $K$-user network that move between nodes with odd indices. In order to achieve fairness between the different users, we can split each message into three parts, apply the above scheme using the first part of each message whose index is not equal to $3$ modulo $4$ (these messages are not transmitted), then shift forward the user indices by one, and apply the same scheme for the next part of each message whose index is not equal to $3$ modulo $4$, and so on. Overall, we obtain an average per user DoF of $\frac{3}{4}$, and each node gets paid as many coins as it paid. To summarize, the \textbf{monetary mechanism} enabled a \textbf{distributed strategy} to reach the \textbf{optimal centralized solution} with a number of coins equal half the number of users in the network. We note that the number of needed coins would grow not only with the number of users, but also with the number of interfering links per receiver. In the illustrated example, each receiver observed only one interfering signal.

\subsection{Transaction Overhead:} In the above example, aside from the coin exchange, each node sends a $2$-bit coordination message to the following node. This incurs an extra communication overhead that may be unjustified if the network topology changes frequently. In other words, the overhead of the coordination and payment protocols and the \textbf{coherence time} of the network topology are factors that determine whether a monetary mechanism leads to overall rate gains. In general, each node would need to send a coordination message to each node whose receiver is connected to the considered node's transmitter, and possibly exchange coins with any neighboring node in the interference conflict graph.

\subsection{Setup Delay:} We note that in Algorithm~\ref{alg:the_alg}, a node generates a coordination message only after receiving a coordination message from the preceding node. Hence, the delay for completion of the coordination protocol would be on the order of the number of users in the network. It would hence be interesting to study an extension of that algorithm where \textbf{local connectivity enables parallel execution} of the algorithm in different parts of the network. 

\section{Conclusion}
In this work, we presented a greedy distributed algorithm that reaches the DoF-optimal centralized solution for linear interference networks, under a backhaul load constraint that allows for distributing each message to an average of one transmitter. The proposed scheme is enabled by a monetary mechanism that has only one coin type, and a coordination protocol that relies on $2$-bit messages. The transmission scheme is based on cooperative zero-forcing and interference avoidance. A node pays a coin either to rent out the payee's transmitter, or because it is causing interference at the payee's receiver. It is interesting to study in future work how the proposed scheme can be extended to apply for general network topologies. In particular, how the number of coins, overhead of the coordination protocol, and the delay scale with the size of the network with an arbitrary topology. The detailed implementation of blockchain techniques to realize the monetary policies advocated in this line of work is also an interesting direction for future research. Finally, an intriguing general direction for future work is to study, from an information theoretic standpoint, how the blockchain could enable new technologies like CoMP, caching, and machine learning to deliver the promise of next generation wireless networks.
\bibliographystyle{IEEEtran}
\bibliography{refs}

\end{document}